# A Systematic Overview of Single-Cell Transcriptomics Databases, their Use cases, and Limitations

**Mahnoor N. Gondal[1,2], Saad Ur Rehman Shah[3], Arul M. Chinnaiyan[1,2,4,5,6,7, †,*], Marcin Cieslik[1,2,4,7, †,\*]**

[1] Department of Computational Medicine & Bioinformatics, University of Michigan, Ann Arbor, MI USA

[2] Michigan Center for Translational Pathology, University of Michigan, Ann Arbor, MI USA

[3] Gies College of Business, University of Illinois Business College, Champaign, IL USA

[4] Department of Pathology, University of Michigan, Ann Arbor, MI USA

[5] Department of Urology, University of Michigan, Ann Arbor, MI USA

[6] Howard Hughes Medical Institute, Ann Arbor, MI USA

[7] University of Michigan Rogel Cancer Center, Ann Arbor, MI USA

**\* Correspondence:**

Marcin Cieslik
mcieslik@med.umich.edu

Arul M. Chinnaiyan
arul@med.umich.edu

**Abstract**

Rapid advancements in high-throughput single-cell RNA-seq (scRNA-seq) technologies and experimental protocols have led to the generation of vast amounts of genomic data that populates several online databases and repositories. Here, we systematically examined large-scale scRNA-seq databases, categorizing them based on their scope and purpose such as general, tissue-specific databases, disease-specific databases, cancer-focused databases, and cell type-focused databases. Next, we discuss the technical and methodological challenges associated with curating large-scale scRNA-seq databases, along with current computational solutions. We argue that understanding scRNA-seq databases, including their limitations and assumptions, is crucial for effectively utilizing this data to make robust discoveries and identify novel biological insights. Furthermore, we propose that bridging the gap between computational and wet lab scientists through user-friendly web-based platforms is needed for democratizing access to single-cell data. These platforms would facilitate interdisciplinary research, enabling researchers from various disciplines to collaborate effectively. This review underscores the importance of leveraging computational approaches to unravel the complexities of single-cell data and offers a promising direction for future research in the field.

# 1 Introduction

The first commercially available single-cell platform emerged in 2014(1). Over the past decade, single-cell sequencing technologies have rapidly advanced, becoming faster and more cost-effective. Today, there are over 10 different commercially available platforms for high-throughput single-cell data collection(2,3). This advancement has fueled remarkable growth in the field of single-cell RNA sequencing (scRNA-seq) research, with nearly 2000 studies published to date(4), populating numerous databases and repositories(5–9). These studies have provided valuable insights into various biological processes, including development(10), disease initiation and progression(11), immune response(12), and identification of rare cell types(13,14). Alongside the generation of large-scale single-cell data, we also observe a sharp rise in scRNA-seq analysis tools, expected to reach 3000 by the end of 2025(15).

Previous reviews and benchmarking analyses have extensively covered various aspects of scRNA-seq analysis such as quality control(16), normalization(17,18), integration(19,20), and cell type annotation(21). However, the complexity of large-scale data necessitates a comprehensive evaluation of available scRNA-seq databases and repositories. This evaluation is crucial to understand concepts like integration in the context of large-scale databasing. Understanding the scope and limitations of these databases is crucial for storing, analyzing, and interpreting single-cell data directly from these repositories. In this review, we systematically address the limitations and common assumptions of existing scRNA-seq databases. We discuss the utility of these databases to meet the specific needs of researchers studying different biological systems and processes.

# 2 Landscape of Single-cell Transcriptomics Databases

The rapid expansion of single-cell RNA-seq (scRNA-seq) studies has led to the development of numerous databases and repositories for storing, retrieving, and interpreting single-cell data(5–9). These databases provide a resource for single-cell transcriptomic data that can be used to build computational models to investigate various biological processes. The data in scRNA-seq databases or atlases can come from either a "primary" source, where the data was generated by the study itself, or an "aggregated" source, where data was collected and curated from multiple studies (**Figure 1A-D, Table 1**). Single-cell databases can be further categorized into general (non-specific or broad category of databases), tissue-specific databases, disease-specific databases, cancer-focused databases, and cell type-focused databases (**Figure 1E**).

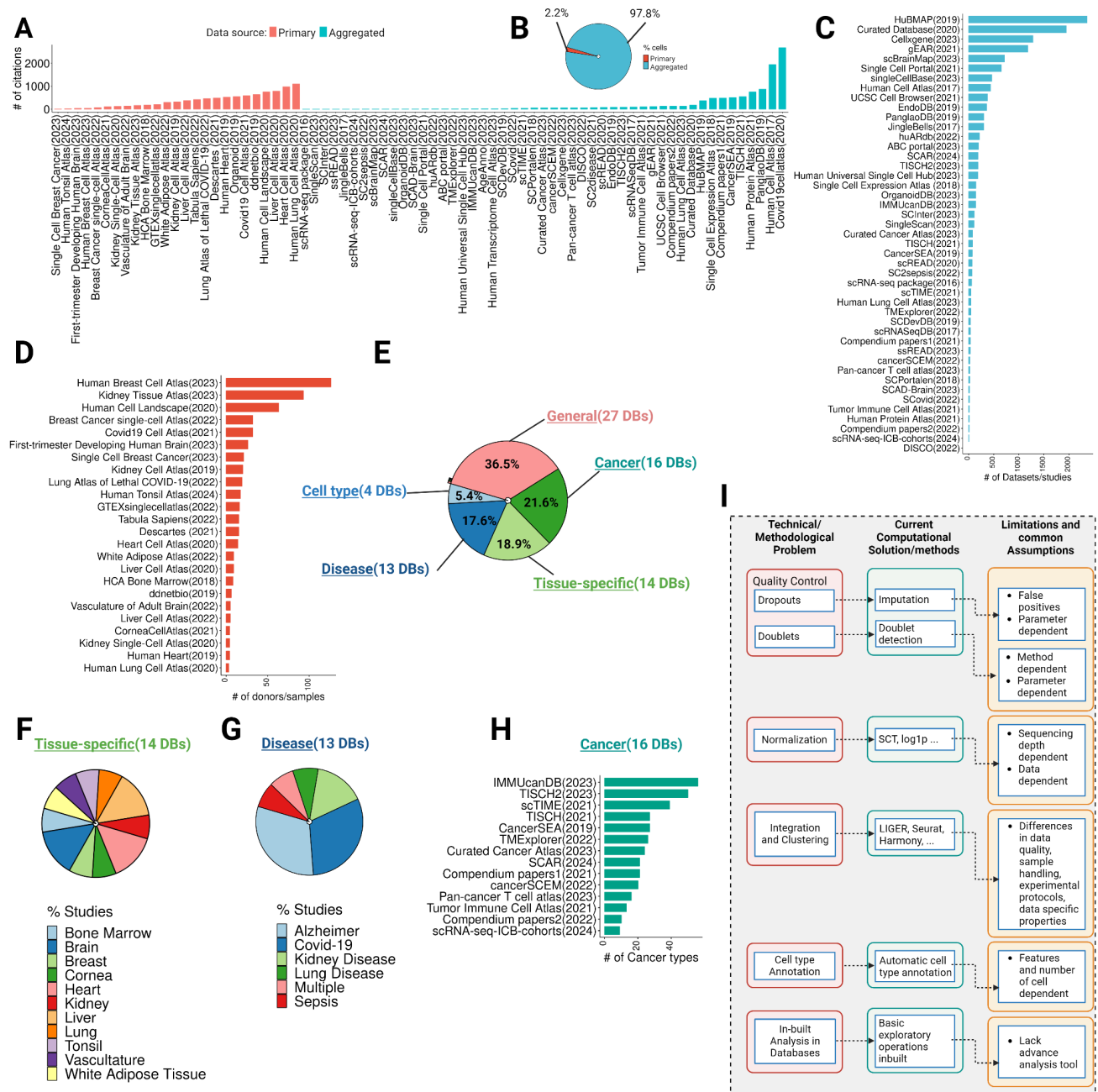

**Figure 1. Overview of single-cell databases, their technical/methodological issues, current solutions, and common assumptions.** (**A**) Overview of citations gathered from single-cell data repositories from primary or aggregated studies (data collected on March 31st). (**B**) The pie chart showing the total cells in the primary and aggregated data source (**C**) Highlights the number of datasets or studies published per database in the primary data source category. (**D**) Shows the number of donors/samples in the aggregated data source category. (**E**) A pie chart highlighting the total percentage of databases in general, tissue-specific, disease-specific, cancer-focused, and cell type-focused databases. (**F-G**) Exhibits the percentage of studies for tissue-specific and disease-focused databases. (**H**) Shows the number of cancer types per cancer-focused databases. (**I**) Technical and methodological issues with databases, current computational and methods-based solutions, and their common assumptions and limitations.

## 2.1 General and Tissue-Specific Single-cell Databases

The establishment of the Human Cell Atlas (HCA)(5) in 2017 marked a significant effort to collect and integrate large-scale single-cell data into a comprehensive reference atlas for all human cells. HCA's open-access resource forms one of the largest public databases for integrated single-cell data from large-scale sequencing projects comprising over 437 projects and 58.5 million cells across 18 tissues. Single-cell atlases offer high-resolution views of cellular composition in organs, leading to groundbreaking discoveries of rare cell types, developmental processes, and cell states associated with various disease processes(22–24). Despite the accessibility and extensive research interest in this data, disagreements may arise when selecting a particular study as a reference for hypothesis building(25).

To address this issue, one of the HCA's sub-projects aims to develop tissue-specific reference atlases that serve as consensus representations of specific organs across multiple projects (**Figure 1F**). These atlases provide a standardized reference for specific tissues for comparing different datasets, facilitating cross-study comparisons and meta-analyses. An example of a tissue-specific single-cell database established by the HCA is the Human Lung Cell Atlas (HLCA)(26) which integrates 49 lung datasets encompassing 2.4 million cells from 486 individuals. The HLCA database development involved four main steps: data curation, integration method selection, cell type annotation, and data usage. Despite the benefits associated with large-scale tissue-specific single-cell databases, it is important to note that integrating data from diverse datasets, labs, and technologies presents challenges due to differences in data quality, sample handling, and experimental protocols. Moreover, ensuring consistency and standardization across datasets is crucial for meaningful comparisons, but achieving this in practice can be complex, particularly with the wide variety of cell types and states. For the HLCA project, the team primarily relied on harmonized manual annotation, integration benchmarking(20), and expert views, which can be subjective and may lack reproducibility.

Other general databases include Single Cell Portal (SCP)(27) and CZ CellxGene Discover(28) which are more flexible and are ideal for retrieving single-cell data focusing on a particular dataset of interest, focusing on unique features and variations within the datasets. The SCP and CZ CellxGene, developed by the Chan Zuckerberg Initiative (CZI) and Broad Institute, respectively, provide web-based interfaces for data exploration and analysis. CZ CellxGene hosts more than 1284 datasets while SCP constitutes 654 datasets. They offer interactive visualization tools for exploring gene expression patterns, cell clusters, and cell type annotations in scRNA-seq data. Both platforms support the sharing of scRNA-seq datasets, allowing researchers to collaborate and access public datasets. Using the CZ CellxGene platform, users can also download the raw count data as an RDS file containing a Seurat object or an h5ad file with an AnnData object to perform their analysis. Similarly, SCP data can be downloaded as individual metadata files, raw count expressions, and normalized expression data from the website directly, however, the availability of raw or normalized data is subjective to each study in SCP. Another interesting example of a comprehensive database is Tabula Sapiens(29) which houses primary data from 15 individuals across 24 tissues. This database enables the evaluation of gene expression in normal or baseline cell states, providing a valuable resource for developing gene regulation networks and trajectories(30). It offers a unique opportunity to study cell type-specific expression changes. The data is easily accessible through web platforms and can also be explored using tools like CZ CellxGene.

2.2 Disease-focused Single-cell Databases

Numerous other scRNA-seq databases have also emerged, including PanglaoDB(9), UCSC Cell Browser(31), and Human Protein Atlas(32). However, these general databases are not designed to systematically gather data on gene expression specificity in different diseases. Given the diverse and heterogeneous nature of human diseases, which manifest unique gene expression profiles, there is a critical need for databases focused on disease-specific exploration (**Figure 1G**).

SC2disease(33) is a manually curated scRNA-seq database that addresses this need, cataloging cell type-specific genes associated with 25 diverse diseases, including Huntington's disease, multiple sclerosis, and Alzheimer's disease. While SC2disease represents a pioneering effort in disease-specific gene expression profiling, there is also a growing need for more specific databases dedicated to the disease of interest. To address this, databases like SC2sepsis(34), ssREAD(35), and SCovid(36) have emerged, focusing on individual diseases such as sepsis, Alzheimer's, and Covid-19, respectively. These databases aim to provide a more granular and disease-specific view of gene expression patterns, enhancing our understanding of disease mechanisms and potential therapeutic targets.

2.3 Cancer-focused Single-cell Databases

Cancer is a complex disease characterized by its highly heterogeneous and multifactorial nature(37). Traditional approaches to studying cancer, such as bulk RNA sequencing constitute a mixture of the cellular composition in tumors and often fail to accurately capture cancer cell-specific gene expression(38,39). scRNA-seq technologies offer unprecedented insights into tumor heterogeneity, evolution, and responses to therapy(40–42). As a result, numerous databases hosting cancer-focused scRNA-seq data have emerged (**Figure 1H**).

One such example is CancerSEA(43) which was launched in 2019 as a resource utilizing single-cell data from cancer datasets to decode the functional states of cancer cells, these states included stemness, invasion, metastasis, proliferation, epithelial-to-mesenchymal transition (EMT), angiogenesis, apoptosis, cell cycle, differentiation, DNA damage, DNA repair, hypoxia, inflammation, and quiescence. In a salient study, Dohmen et al.(44) utilized CancerSEA's functional states to validate gene sets derived from their machine-learning model, while Zhao et al.(45) demonstrated the necessity of NF-KB for initiating oncogenesis using CancerSEA's functional states. Several other studies(46–49) have leveraged CancerSEA to correlate their gene or gene set findings with cancer single-cell data, showcasing the utility of this resource.

However, CancerSEA has limitations, including hosting only 93,475 malignant cells and an inability to study interactions between stromal or immune cells and cancer cells. It also lacks a user-friendly web interface to support data exploration and visualization. In an attempt to overcome these challenges, TISCH was originally developed in 2021(50), with version 2 released in 2023(51). TISCH2 curates cancer datasets with both malignant and non-malignant cell types, currently hosting

190 datasets, encompassing 50 cancer types, and spanning 6 million cells. To illustrate the utility of TISCH in computational models, Xu et al.(52) employed TISCH's data to analyze the correlation between FOXM1 and immune cells. Similarly, Zhang et al.(53) employed TISCH to evaluate m7G regulators expression in osteosarcoma scRNA-seq data. As such, numerous studies have utilized TISCH to evaluate the expression of genes of interest across cancer datasets(53–58). Although TISCH provides a valuable resource to the cancer research community, it is important to be aware of the assumptions and limitations of TISCH data. While many studies aim to understand gene expression in malignant cells, TISCH also contains treatment data from immune checkpoint blockade (ICB), chemotherapy, and targeted therapy. Therefore, it is important to ensure that the results are not confounded, as gene expression varies after treatments and can yield diverse results(59). Additionally, TISCH includes data from multiple stages of cancer, such as primary tumors or metastatic sites. Therefore, users need to carefully extract only relevant information when employing TISCH data. TISCH employs an automatic cell-type annotation method, which may lead to a lack of consensus with the original dataset's manual annotation. Importantly, all downloaded datasets in TISCH are in fixed expression matrices(41), and users cannot download the raw count data. Therefore, any attempts to further integrate or normalize the data might result in technical variation rather than biological results. These limitations could potentially introduce bias into analyses or hinder comparability across different databases.

2.4 Cell-type-focused Single-cell Databases

To better understand the intricacies of cell biology, dedicated resources focused on cell-type profiling of single cells have emerged. JingleBells(60), introduced in 2017, represented an advancement in this direction by providing a comprehensive immune cell resource. JingleBells facilitates the study of immune cell involvement in various diseases, including cancer, and infectious diseases, providing valuable insights into disease mechanisms and potential therapeutic targets. However, JingleBells lacks an interactive web interface and only allows for BAM file download which means analyzing and interpreting single-cell data from JingleBells requires specialized computational tools and expertise, limiting accessibility to researchers with specific skills. In comparison, the human Antigen Receptor database (huARdb)(61), published originally in 2022, is a comprehensive human single-cell immune profiling database, housing 444,794 high-confidence T or B cells (hcT/B cells) with complete TCR/BCR sequences and transcriptomes sourced from 215 datasets. To enhance user experience, the authors have created a user-friendly web interface that offers interactive visualization modules, enabling biologists to analyze transcriptome and TCR/BCR features at the single-cell level with ease. Fan et al.(62) utilized huARdb by analyzing ulcerative colitis (UC) patients' immune cells derived from huARdb. Similarly, they also employed huARdb to investigate the healthy and UC composition of peripheral blood immune cells and colonic cells(63). Additional cell-type-focused single-cell databases include EndoDB(64) which hosts endothelial cells transcriptomics data from 360 datasets and ABC portal(65), a database for blood cells across 198 datasets, allowing for a blood cell-type-specific exploration.

| # | Database | Year | PMID | Citations | URL | Web Interference | Number of studies/dataset | Number of donors/samples | Number of cells | Data source | Cancer types/Diseases | Number of tissues | Number of organs | Specific focus | Tissue type | Download available |
|---|---|---|---|---|---|---|---|---|---|---|---|---|---|---|---|---|
| 13 | Developing Human Brain | 2023 | 37624600 | 30 | https://hdca-sweden.scilifelab.se/tissues-overview/brain | Available | 1 | 20 | 100837 | Primary | | 1 | | Tissue-specific | Brain | Available |
| 14 | HCA Bone Marrow | 2018 | 30243574 | 179 | https://www.altanalyze.org/ICGS/HCA/splash.php | Available | 1 | 8 | 100000 | Primary | | 1 | | Tissue-specific | Bone Marrow | Not available |
| 15 | Human Cell Landscape | 2020 | 32214235 | 750 | https://bis.zju.edu.cn/HCL/ | Available | 1 | 63 | 700,000 | Primary | | 50 | | General | General | Available |
| 16 | Descartes | 2021 | 33184181 | 486 | https://descartes.brotmanbaty.org/ | Available | 1 | 15 | 4000000 | Primary | | 121 | 15 | General | General | Available |
| 17 | Tabula Sapiens | 2022 | 35549404 | 417 | https://tabula-sapiens-portal.ds.czbiohub.org/ | Available | 1 | 15 | 500,000 | Primary | | 24 | | General | General | Available |
| 18 | GTEX singlecell atlas | 2022 | 35549429 | 201 | https://www.gtexportal.org/home/singleCellOverviewPage | Available | 1 | 16 | 209126 | Primary | | 8 | | General | General | Available |
| 19 | Kidney Single-Cell Atlas | 2020 | 32978267 | 121 | http://www.ruuo-kidney-gene-atlas.com | Available | 1 | 4 | 17136 | Primary | | 1 | | Disease | Kidney Disease | Not available |
| 20 | Kidney Tissue Atlas | 2023 | 37468583 | 161 | https://cellxgene.cziscience.com/collections/bcb61471-2a44-4d00-a0af-ff085512674c | Available | 1 | 93 | 584,843 | Primary | | 1 | | Disease | Kidney Disease | Available |
| 21 | Covid19 Cell Atlas | 2021 | 33915569 | 586 | https://singlecell.broadinstitute.org/single_cell/study/SCP1052/covid-19-lung-autopsy-samples#study-summary | Available | 1 | 32 | 106792 | Primary | | 11 | | Disease | Covid-19 | Available |
| 22 | Lung Atlas of Lethal COVID-19 | 2022 | 33915568 | 458 | https://singlecell.broadinstitute.org/single_cell/study/SCP1219/columbia-university-nyp-covid-19-lung-atlas | Available | 1 | 19 | 116314 | Primary | | 1 | | Disease | Covid-19 | Available |
| 23 | ddnetbio | 2019 | 31768052 | 636 | http://adsn.ddnetbio.com/ | Available | 1 | 6 | 13214 | Primary | | 1 | | Disease | Alzheimer | Available |
| 24 | Breast Cancer single-cell Atlas | 2022 | 35361816 | 65 | https://bcatlas.tigem.it/tigem/dibernardo/AIRC_atlas_32_ccls/?ds=Atlas_32_ccls | Available | 1 | 32 | 35276 | Primary | 1 | 1 | | Cancer | Breast Cancer | Available |
| 25 | Single Cell Breast Cancer | 2023 | 37158690 | 4 | https://mikagiao.shinyapps.io/scBC/ | Available | 1 | 21 | 117958 | Primary | 1 | 1 | | Cancer | Breast Cancer | Not available |
| 26 | scRNA-seq package | 2016 | NA | 0 | https://bioconductor.org/packages/release/data/experiment/vignettes/scRNAseq/inst/doc/scRNAseq.html | Not available | 61 | | 1641896 | Aggregated | | | | General | General | Available |
| 27 | Human Cell Atlas | 2017 | 29206104 | 1942 | https://data.humancellatlas.org | Available | 437 | 8600 | 58500000 | Aggregated | | | 18 | General | General | Available |
| 28 | scRNASeqDB | 2017 | 29206167 | 91 | https://bioinfo.uth.edu/scrnaseqdb | Available | 38 | 13,440 | 13440 | Aggregated | | | | General | General | Not available |
| 29 | Single Cell Expression Atlas | 2018 | 31665515 | 475 | https://www.ebi.ac.uk/gxa/sc/home | Available | 147 | | 10,505,726 | Aggregated | | | | General | General | Available |
| 30 | SCPortalen | 2018 | 29045713 | 51 | http://single-cell.clst.riken.jp/ | Available | 23 | | 20,761 | Aggregated | | | | General | General | Available |
| 31 | SCDevDB | 2019 | 31611909 | 20 | https://scdevdb.deepomics.org | Available | 38 | | 13,440 | Aggregated | | | | General | General | Not available |
| 32 | PanglaoDB | 2019 | 30951143 | 867 | https://panglaodb.se/ | Available | 305 | | 1,126,580 | Aggregated | | 74 | | General | General | Available |
| 33 | HuBMAP | 2019 | 31597973 | 368 | https://portal.hubmapconsortium.org/ | Available | 2362 | 1843 | | Aggregated | | 31 | | General | General | Available |
| 34 | Curated Database | 2020 | 33247933 | 174 | https://docs.google.com/spreadsheets/d/1En7-UV0k0IaDilfjFkdn7dggyR7jIk3WH8QgXaMOZF0/edit#gid=0 | Not available | 1946 | | 134393568 | Aggregated | | | | General | General | Not available |
| 35 | Human Protein Atlas | 2021 | 34321199 | 749 | https://www.proteinatlas.org/humanproteome/single+cell+type | Available | 14 | | 174271 | Aggregated | | 13 | 14+ | General | General | Available |
| 36 | Single Cell Portal | 2021 | 37502904 | 8 | https://singlecell.broadinstitute.org/single_cell | Available | 654 | | 40,699,488 | Aggregated | | | | General | General | Available |
| 37 | UCSC Cell Browser | 2021 | 34244710 | 127 | https://cells.ucsc.edu/? | Available | 378 | | | Aggregated | | | | General | General | Available |
| 38 | gEAR | 2021 | 34172972 | 112 | https://www.umgear.org | Available | 1180 | | | Aggregated | | | | General | General | Not available |
| 39 | DISCO | 2022 | 34791375 | 55 | https://www.immunesinglecell.org/ | Available | 1,077 | 13,996 | 61,280,618 | Aggregated | 393 | 174 | | General | General | Available |
| 40 | singleCellBase | 2023 | 37730627 | 3 | http://cloud.capitalbiotech.com/SingleCellBase/ | Available | 464 | | 9,156 | Aggregated | 464 | 165 | | General | General | Not available |
| 41 | SingleScan | 2023 | 38062357 | 0 | http://cailab.labshare.cn/SingleScan | Available | 109 | | 3077622 | Aggregated | 49 | 5+ | | General | General | Available |
| 42 | AgeAnno | 2023 | 36200638 | 14 | https://relab.xidian.edu.cn/AgeAnno/#/ | Available | | 226 | | Aggregated | 28 | 17 | | General | General | Available |
| 43 | scBrainMap | 2023 | 37195696 | 2 | https://scbrainmap.sysneuro.net/ | Available | 715 | | 6577222 | Aggregated | 20 | 1 | | General | General | Available |
| 44 | Human Transcriptome Cell Atlas | 2023 | 36130266 | 14 | https://www.htcatlas.org/ | Available | | 19 | 24652615 | Aggregated | | 19 | | General | General | Available |
| 45 | Human Universal Single Cell Hub | 2023 | 36318258 | 10 | http://husch.comp-genomics.org/#/search | Available | 185 | | 300000 | Aggregated | | 45 | | General | General | Available |
| 46 | CellxGene | 2023 | https://doi.org/10.1101/2021.04.05.436318 | 54 | https://cellxgene.cziscience.com/ | Available | 1284 | | 85100000 | Aggregated | | | | General | General | Available |
| 47 | SCInter | 2023 | 38125297 | 0 | https://bio.liclab.net/SCInter/index.php | Available | 115 | 1016 | | Aggregated | | | | General | General | Available |
| 48 | OrganoidDB | 2023 | 36271792 | 4 | http://www.inbirg.com/organoid_db/ | Available | 145 | | 670000 | Aggregated | | 141 | | General | General | Not available |
| 49 | SC2sepsis | 2022 | 35980266 | 1 | http://www.rjh-sc2sepsis.com/index | Available | 71 | | 232226 | Aggregated | 1 | | | Disease | Sepsis | Available |
| 50 | SC2disease | 2021 | 33010177 | 67 | http://easybioai.com/sc2disease/ | Available | | | 946481 | Aggregated | 25 | 29 | | Disease | Multiple | Available |
| 51 | Human Lung Cell Atlas | 2023 | 37291214 | 130 | https://cellxgene.cziscience.com/collections/6f6d381a-7701-4781-935c-db10d30de293 | Available | 49 | 486 | 2,400,000 | Aggregated | 15 | 4 | | Disease | Lung Disease | Available |
| 52 | Covid19cellatlas | 2020 | 32327758 | 2676 | https://www.covid19cellatlas.org | Available | | 15 | | Aggregated | 1 | 29 | | Disease | Covid-19 | Available |
| 53 | SCovid | 2022 | 34634820 | 30 | http://bio-annotation.cn/scovid | Not available | 21 | | 1042227 | Aggregated | 1 | 10 | | Disease | Covid-19 | Not available |
| 54 | scREAD | 2020 | 33241205 | 74 | https://bmblx.bmi.osumc.edu/scread/ | Available | 73 | | 713,640 | Aggregated | 1 | | | Disease | Alzheimer | Available |
| 55 | ssREAD | 2023 | 37745592 | 0 | https://bmblx.bmi.osumc.edu/ssread/ | Available | 35 | | 2,572,355 | Aggregated | 1 | | | Disease | Alzheimer | Available |
| 56 | SCAD-Brain | 2023 | 37251804 | 4 | https://www.bioinform.cn/SCAD/ | Available | 21 | 359 | 1,564,825 | Aggregated | 1 | 10 | | Disease | Alzheimer | Not available |
| 57 | JingleBells | 2017 | 28416714 | 0 | https://jinglebells.bgu.ac.il/ | Available | 302 | | | Aggregated | | | | Cell type | NA | Available |
| 58 | EndoDB | 2019 | 30357379 | 78 | https://endotheliomics.shinyapps.io/endodb/ | Available | 360 | | 5847 | Aggregated | | 15+ | | Cell type | NA | Available |
| 59 | huARdb | 2022 | 34606616 | 9 | https://huarc.net/v2/ | Available | 215 | | 444794 | Aggregated | 12 | 24 | | Cell type | NA | Not available |
| 60 | ABC portal | 2022 | 35920330 | 9 | http://abc.sklehabc.com/#/home | Available | 198 | | 3878681 | Aggregated | 59 | 10+ | | Cell type | NA | Available |
| 61 | CancerSEA | 2019 | 30329142 | 505 | http://biocc.hrbmu.edu.cn/CancerSEA/ | Available | 74 | | 93,475 | Aggregated | 27 | | | Cancer | Multiple | Available |
| 62 | scTIME | 2021 | 34117085 | 33 | http://scTIME.sklehabc.com | Available | 49 | | 196273 | Aggregated | 39 | | | Cancer | Multiple | Available |
| 63 | TISCH | 2021 | 33179754 | 542 | http://tisch1.comp-genomics.org/ | Available | 76 | | 2,045,746 | Aggregated | 27 | | | Cancer | Multiple | Available |
| 64 | Compendium papers1 | 2021 | 34914499 | 485 | http://cancer-pku.cn:3838/PanC_T/ | Available | 37 | 316 | 397,810 | Aggregated | 21 | | | Cancer | Multiple | Available |
| 65 | Tumor Immune Cell Atlas | 2021 | 34548323 | 99 | https://singlecellgenomics-cnag-crg.shinyapps.io/TICA/ | Available | 16 | 217 | 500,000 | Aggregated | 13 | | | Cancer | Multiple | Available |
| 66 | TMExplorer | 2022 | 36084081 | 9 | https://figshare.com/projects/TMExplorer_A_Tumour_Microenvironment_Single-cell_RNAseq_Database_and_Search_Tool/101471 | Available | 48 | | 1438299 | Aggregated | 26 | | | Cancer | Multiple | Available |
| 67 | cancerSCEM | 2022 | 34643725 | 52 | https://ngdc.cncb.ac.cn/cancerscem | Available | 28 | | 638,341 | Aggregated | 20 | | | Cancer | Multiple | Available |
| 68 | Compendium papers2 | 2022 | 36333338 | 129 | https://gist-fgl.github.io/sc-caf-atlas/ | Available | 12 | 226 | 855,271 | Aggregated | 10 | 10 | | Cancer | Multiple | Available |
| 69 | IMMUcanDB | 2023 | 36459564 | 12 | https://immucandb.vital-it.ch/ | Available | 144 | | 4474385 | Aggregated | 56 | | | Cancer | Multiple | Available |
| 70 | TISCH2 | 2023 | 36321662 | 82 | http://tisch.comp-genomics.org/ | Available | 190 | | 6,297,320 | Aggregated | 50 | | | Cancer | Multiple | Available |
| 71 | Curated Cancer Atlas | 2023 | 37258682 | 51 | https://www.weizmann.ac.il/sites/3CA/ | Available | 77 | | 2,591,545 | Aggregated | 24 | | | Cancer | Multiple | Available |
| 72 | Pan-cancer T cell atlas | 2023 | 37248301 | 54 | https://singlecell.mdanderson.org/ | Available | 27 | 670 | 656,742 | Aggregated | 16 | 21 | | Cancer | Multiple | Not available |
| 73 | SCAR | 2024 | 37739405 | 2 | http://8.142.154.29/SCAR2023/ | Available | 190 | | 11301352 | Aggregated | 21 | 395 | | Cancer | Multiple | Not available |
| 74 | scRNA-seq ICB cohorts | 2024 | 38328153 | 0 | https://zenodo.org/records/10407126 | Available | 8 | | 90270 | Aggregated | 9 | | | Cancer | Multiple | Available |

**Table 1. Detailed list of the single-cell databases.** The list encompasses essential information such as database name, year of establishment, PubMed ID (PMID), number of citations, URL, web interface availability, number of datasets/studies per database, cell count, primary vs aggregated distinction, specific groups, tissue type, and download availability.

# 3 Challenges Associated with the Utilization of Large-scale Single-cell Databases and their Examples from Current Literature

While the scRNA-seq field is progressing towards the improvement and development of large-scale single-cell databases, their application in research comes with certain caveats and despite their vastness, they must be used judiciously (**Figure 1I**). Some of the key considerations and limitations include:

## 3.1 Data quality

Ensuring data quality in scRNA-seq is critical for accurate interpretation and analysis. A fundamental assumption of droplet-based scRNA-seq is that each droplet, where molecular tagging and reverse transcription occur, contains messenger RNA (mRNA) from a single cell. However, in practice, this assumption is often violated, leading to potential distortions in the interpretation of scRNA-seq data. Common examples include droplets containing multiple cells (doublets) or no cells at all (empty droplets or dropouts). This becomes a major issue because large-scale databases such as the Human Cell Atlas (HCA) rely heavily on the accuracy and cellular specificity of transcriptional readouts generated by scRNA-seq.

### 3.1.1 Examples from current literature and benchmarking studies

**Dropouts:** To overcome the issue of dropouts, numerous single-cell imputation methods have been developed. However, imputation affects downstream results and some of these methods may introduce false correlations. For example, Breda et al.(66)'s comparison of MAGIC(67) results with Sanity (SAmpling-Noise-corrected Inference of Transcription activitY), elicited that MAGIC introduced strong positive correlations where no or low correlation was expected. A comparative study by Zhang et al.(68) highlighted that the number of cells and method parameters also affected imputation results and some methods preferred similar cells while imputing. Therefore, imputation results can be variable, and downstream analysis will be affected by imputation therefore in our opinion it should ideally be avoided or performed with caution.

**Doublets:** Doublets are also not real cells and are major confounders in scRNA-seq data analysis. However, there are computational methods that exist to detect doublets in single-cell data. A benchmarking study(69) compared nine doublet detection methods, revealing that there is still room for improvement in detection accuracy. Generally, these methods performed better on datasets with higher doublet rates, larger sequencing depths, more cell types, or greater heterogeneity between cell types. However, the removal of doublets by these methods led to improvements in various downstream analyses. It enhanced the identification of Differentially Expressed (DE) genes, reduced the presence of spurious cell clusters, and improved the inference of cell trajectories. However, the extent of improvement varied across different methods, highlighting the need for further refinement and development in this area.

## 3.2 Normalization

Normalization is another critical aspect of scRNA-seq data analysis and can be a complex problem when dealing with multiple datasets. Specifically, variability in experimental protocols and data processing methods can pose challenges in data normalization, affecting the comparability of results across datasets in a database. Differences in normalization approaches can lead to discrepancies in gene expression profiles, making it difficult to draw meaningful conclusions from the downstream analyses.

3.2.1 Examples from current literature and benchmarking studies

There are several methods to perform single-cell data normalization such as SCT transformation, and log1p normalization(17). The choice of the method, however, is dependent on various features of the data including sequencing depth as both lowly and highly abundance genes are confounded by sequencing depth(17). Booeshaghi et al.(18) demonstrated that the assumptions implied in the choice of normalization methods will affect downstream analysis in determining whether the variation is technical or biological. In a salient example, TISCH2(51) database hosts single-cell gene expression matrices for each dataset. In our analysis of TISCH2 data, this matrix is already normalized and integrated, users incorporating this data in their research need to be aware of this normalization to make accurate assessments of data and not re-normalize or merge it directly with other datasets which might result in substandard results. Therefore, in our opinion, when using datasets directly from single-cell databases it is necessary to be aware of the pre-processing steps and how they affect downstream results to ensure accurate analysis and interpretation.

3.3   Integration and batch effects removal

The integration and batch effect removal of scRNA-seq data from diverse datasets, labs, and technologies can be complex(70). Variations in data formats, processing pipelines, and batch effects can affect the robustness and reliability of integrated analyses, potentially masking true biological signals. Methods for integrating heterogeneous datasets are continually evolving, with efforts focused on minimizing batch effects and preserving biological variability. There are more than 50 integration methods published to date(20,71).

3.3.1 Examples from current literature and benchmarking studies

Large databases host numerous datasets from multiple studies, however, it is also important to be aware of the properties associated with each study during integration. For example, Salcher et al.(13) established a large non-small cell lung cancer (NSCLC) atlas comprising 29 datasets spanning 1,283,972 single cells from 556 samples. Although this effort resulted in the in-depth characterization of a neutrophil subpopulation, however, according to our re-analysis of this data, among the 29 datasets, Maynard et al.(72)'s NSCLC samples were also incorporated which were not treatment-naive. This can be a potential confounder in downstream analysis. Therefore, it is the user's responsibility to be aware of this data-specific property and to use atlases and databases with care to derive robust biological insights. Similarly, several attempts have been made to benchmark integration methods for single-cell data(19,20). While Tran et al.(19) showed that LIGER(73), Seurat 3(74), and Harmony(75) performed the best among 11 other methods, Luecken et al.(20) revealed that LIGER and Seurat v3 favor the removal of batch effect over the conservation of biological variation. This highlights the importance of considering the dataset and the specific research question when selecting an integration method. In our view, similar to the Human Cell Atlas pipeline, benchmarking integration methods need to be performed on each study to first evaluate which method suits your data the best. Selecting the right method is crucial as it directly impacts the biological insights that can be generated from the integrated data.

3.4 Cell-Type Annotation

Accurate annotation of cell types in scRNA-seq databases is crucial for interpreting results accurately. While automatic cell-type annotation methods are convenient, they may lack consensus with manual annotations from original datasets. This can introduce ambiguity in cell-type assignments and lead to misinterpretation of results. Although harmonizing cell type annotations across different datasets is essential for facilitating cross-study comparisons and meta-analysis, in our opinion the results should be manually validated to make sure the automatic annotations make logical sense.

3.4.1 Examples from current literature and benchmarking studies

In our recent re-analysis of Tabula Sapiens data(29), we observed that 10% of the heart cells were mislabelled as hepatocytes in the study's original metadata. This is biologically incorrect since hepatocytes cannot be in the heart, these are liver epithelial cells(76). One potential reason for this mislabelling can be that Tabula Sapiens data was annotated using an automatic cell-type annotation tool, another reason could be sample mishandling. Therefore, diligent manual intervention for cell type annotation needs to be practiced to ensure accurate and robust results. Additionally, Abdelaal et al.(21) carried out a performance comparison analysis between 22 automatic cell-type identification methods in single-cell data. Although the authors did not state a preference they noted that the results can vary depending on input features and the number of cells which means that they cannot be solely relied on, there will be some manual intervention for accurate cell type annotation.

3.5 "Zero-code" Single-cell Analysis Platforms

Single-cell data plays a crucial role in validating and enhancing the accuracy of wet lab results and hypothesis-driven publications(77). To facilitate easy access and analysis of this data, many databases provide built-in tools that allow researchers without computational expertise to explore existing datasets and assess their hypotheses using basic operations like exploring gene expressions, isolating cell subsets for individual analysis, and identifying clusters within the data. However, for more complex analyses that require significant computational resources, these tools are often not available directly on the database platforms.

To address this challenge, wet lab scientists can employ platforms like ICARUS_v3(78) for zero-code single-cell analysis. ICARUS_v3 employs a geometric cell sketching method to subsample representative cells from the dataset to store in memory. This enables advanced scRNA-seq analysis through a user-friendly web interface. ICARUS_v3 can seamlessly integrate with output files from databases like Single Cell Portal (SCP)(27) and CZ CellxGene Discover(28), eliminating the need for coding expertise. Users can leverage this platform to conduct a wide range of analyses, including differential expression analysis, gene regulatory network construction, trajectory analysis, and cell-cell communication inference. While ICARUS_v3 has its assumptions and limitations, it offers users the flexibility to set parameters at each stage and provides a diverse range of operations. This capability not only bridges the gap between experimental and bioinformatics researchers but also simplifies the utilization of scRNA-seq databases.

## 4 Platforms for hosting and visualizing large-scale single-cell data

As the volume of single-cell data continues to grow, scalability becomes a significant concern. Developing methods and infrastructure that can handle the increasing complexity and size of single-cell datasets is crucial for future research. Towards this aim, for easy, fast, and customizable exploration of single-cell data for public use, numerous user-driven platforms have emerged(79,80).

One such platform, the Interactive SummarizedExperiment Explorer (iSEE)(79), launched in 2018, enables users to host their SummarizedExperiment data. Researchers such as Graf et al.(81) and Newton et al.(82) have employed iSEE to visualize their single-cell data, demonstrating its utility in data exploration. Similarly, the Single Cell Explorer(80) allows users to input loom and Seurat objects, making the data more accessible.

ShinyCell(83) is another example of a platform offering web-based interfaces for exploring and analyzing data. These interfaces can be customized for maximum usability and can be uploaded to online platforms to broaden access to published data. ShinyCell supports various common single-cell data formats, including SingleCellExperiment, h5ad, loom, and Seurat objects, as inputs. In a salient example, Ma et al.(84) used ShinyCell to host their pan-cancer single-cell data, showcasing its versatility and effectiveness in data dissemination. Likewise, Curras-Alonso et al.(85) developed their web application using ShinyCell, highlighting its widespread adoption in the research community. By providing easy-to-use tools for data analysis, these platforms help democratize access to single-cell data and facilitate collaboration between researchers from different disciplines.

## 5 Discussion

The rapid expansion of single-cell RNA-seq (scRNA-seq) studies has ushered in a plethora of databases and repositories dedicated to storing, retrieving, and interpreting single-cell data. These databases provide a wealth of single-cell transcriptomic data that can be used to build computational models to understand various biological processes. However, challenges such as data quality, normalization, integration, and annotation can affect the reliability and comparability of results across different datasets and studies.

While the existing databases are valuable for basic scRNA-seq analysis, they cannot often perform advanced analyses such as regulon activity assessment, pseudobulking, and differential gene expression analysis. Users still need to possess programming skills and be familiar with using a command-line interface to conduct customized analysis. Furthermore, many wet labs may not have the necessary resources to manage high-performance computing clusters. To address this gap and enable wet-lab researchers to conduct advanced scRNA-seq analysis, platforms like ICARUS_v3(78) offer web-based analysis tools. These platforms provide an accessible way for researchers to explore and analyze single-cell data, bridging the gap between wet lab experimentation and bioinformatics analysis.

Taken together, in this mini-review, we address the utility and applicability of large-scale scRNA-seq databases. We address some of the challenges and common assumptions that need to be considered when using these databases for hypothesis-driven studies, highlighting platforms for hosting customized scRNA-seq data for community usage. While challenges remain, the development of user-friendly platforms is narrowing the gap between wet-lab experimentation and bioinformatics analysis, ultimately advancing our understanding of cellular processes at a single-cell level.

# 6 Acknowledgments

The study was supported by the National Cancer Institute (NCI) Outstanding Investigator Award R35CA231996 (A.M.C.), NCI Prostate SPORE grant P50CA186786 (A.M.C.), and NCI Michigan-VUMC Biomarker Characterization Center grant U2CCA271854 (A.M.C.). A.M.C. is also a Howard Hughes Medical Institute Investigator, A. Alfred Taubman Scholar, and American Cancer Society Professor. This manuscript was also supported in part by funding from the Innovation in Cancer Informatics (ICI398672) and the V Foundation for Cancer Research (T2019-006) to M.C. We would like to acknowledge the use of ChatGPT, an AI language model developed by OpenAI, for assisting with language editing and suggestions during the preparation of this manuscript.

# 7 Author contributions

MC, AMC, SURS, and MNG carried out the study design and drafted the manuscript. All authors contributed to the article and approved the submitted version.

# 8 Competing interests

A.M.C. is a co-founder of and serves as a Scientific Advisory Board member for LynxDx, Esanik Therapeutics, Medsyn, and Flamingo Therapeutics. A.M.C. is a scientific advisor or consultant for EdenRoc, Aurigene Oncology, Ascentage Pharma, Proteovant, Belharra, Rappta Therapeutics, and Tempus.